\documentclass{article}
\usepackage[dutch,british]{babel}
\usepackage{graphicx,amsfonts,amsthm,bm,stmaryrd,hyperref,caption,subcaption,amsmath,xurl,hyphenat}

\usepackage[numbers,square,comma]{natbib}
\newtheorem{example}{Example}
\theoremstyle{definition}
\newtheorem{definition}{Definition}

\newcommand{\cP}{\mathcal{P}}
\newcommand{\bR}{\mathbb{R}}
\newcommand{\bN}{\mathbb{N}}
\newcommand{\cN}{\mathcal{N}}

\newcommand{\tup}[1]{\bm{#1}}
\newcommand{\rl}[1]{\textsc{#1}}
\newcommand{\app}[1]{\left\llbracket #1 \right\rrbracket}

\DeclareMathOperator{\cost}{cost}

\title{An Empirical Analysis of Participatory Budgeting in Amsterdam}
\author{Pelle $\mathbb{N}$elissen\footnote{The paper is a MSc Logic student project by the author under supervision of Ulle Endriss, Jan Maly and Simon Rey at the Institute for Logic, Language and Computation (ILLC) of the University of Amsterdam.}}
\date{
Institute for Logic, Language and Computation (ILLC),\\
University of Amsterdam\\[2ex]
\today
}

\begin{document}

\maketitle

\section*{Abstract}
Using data from 35 Participatory Budgeting instances in Amsterdam, we empirically compare two different Participatory Budgeting rules: the greedy cost welfare rule and the Method of Equal Shares. We quantify how proportional, equal and fair the rules are and conclude that, for a small price in total voter satisfaction, the Method of Equal Shares performs better on all notions of fairness studied. We further provide a popular and a visual explanation of the Method of Equal Shares.

\section{Introduction}
Participatory Budgeting (PB) is the practice of democratically distributing a budget over proposed projects. \citet{rey2023computational} published a comprehensive survey of the literature studying and designing different incarnations of PB. Common to most PB processes is the collection of projects proposed by citizens, a voting phase for all citizens and the consequent allocation of funds to a subset of all proposed projects. However, the exact method of collecting projects, organising the voting phase and mapping the votes to a budget allocation, varies. In this project, we will focus on the latter: how does one map votes to a `fair' budget allocation? In particular, we will empirically study the performance of two such Participatory Budgeting rules: the greedy cost welfare rule and the Method of Equal Shares.

The data was collected from a number of PB instances organised by the municipality of Amsterdam. Since 2019, the municipality organises yearly PB instances in different districts under the title ``Buurtbudget'' (``Neighbourhood budget'') or ``Amsterdam begroot'' (``Amsterdam budgets''). In the actual PB processes, a version of the greedy cost welfare rule was used to allocate the available budget. The aim of this project is to compare the budget allocation under the greedy cost welfare rule with the allocation that the Method of Equal Shares would have yielded.

The paper is organised as follows. In Section~\ref{sec:PB}, we briefly introduce the relevant notions from the field of PB. In Section~\ref{sec:results}, we present the results of the analysis. We draw conclusions and formulate recommendations for authorities organising a PB instance in Section~\ref{sec:conclusion}. In the appendix, we provide a short explanation of the Method of Equal Shares for the general public.

\section{Participatory Budgeting}
\label{sec:PB}

In this section, we provide definitions of different notions in PB for reference and notation, but we refer to \citet{rey2023computational} for a more extensive introduction to PB. In the rest of the text, by PB we always refer to \emph{indivisible} participatory budgeting, i.e., each project is either fully funded or it receives no funding; we cannot partially fund the cost of a project.

\subsection{Basic Definitions}

Consider a fixed set of voters $\cN = \{1,\dots, n\}$.  A PB \emph{instance} $I = (\cP, c, b)$ consists of a finite set of \emph{projects} $\cP = \{p_1,\dots,p_m\}$, a \emph{cost function} $c\colon\cP\to\bR_{>0}$ and a \emph{budget limit} $b\in\bR_{>0}$. An \emph{approval ballot} for voter $i\in\cN$ is a map $A_i\colon\cP\to\{0,1\}$ indicating which projects voter $i$ approves of (where $i$ \emph{approves of} $p\in\cP$ if $A_i(p) = 1$). A \emph{profile (of approval ballots)} $\tup{A} = (A_1,\dots,A_n)$ is tuple of (approval) ballots $A_i$ for each voter $i\in\cN$.

Given a PB instance $I = (\cP, c, b)$, a PB rule is a function $R$ that maps possible profiles to subsets of $\cP$. We say that a PB rule $R$ \emph{selects} a project $p\in\cP$ under profile $\tup{A}$ if $p\in R(\tup{A})$. Perhaps the simplest rule to decide on a budget allocation, given some profile of approval ballots, is to rank the projects by the number of approvers and fund projects starting from the top of the list (only skipping a project if its cost exceeds the remaining budget). In fact, this procedure yields a decision rule that approximates maximising the total welfare of all voters, if the welfare of a voter is proportional to the total cost of selected projects that she approves of \citep{talmon2019framework}. Therefore, the rule is referred to as the \emph{greedy cost welfare rule} \rl{GreedCost}.

\begin{definition}[Greedy Cost Welfare Rule]
    The \emph{greedy cost welfare rule} \rl{GreedCost} is the PB rule generated by the following process. Given a set of voters~$\cN$, a PB instance~$I = (\cP, c, b)$ and a profile $\tup{A}$, assign to each project $p\in\cP$ the score $\sigma(p):= \sum_{i\in\cN} A_i(p)$. Construct a subset $\pi$ of $\cP$ as follows. Initially, $\pi = \emptyset$. Iteratively, consider the project $p\in\cP$ with the highest score $\sigma(p)$ that has not yet been considered (if multiple such projects exist, apply a given tie-breaking rule). If $c(\pi\cup\{p\})\leq b$, add $p$ to $\pi$. Once all projects have been considered, return $\pi$.
\end{definition}

Although the assumption that the welfare of individual voters is proportional to the total cost of selected approved projects is certainly debatable, the choice for \rl{GreedCost} can (under this assumption) be justified by arguing that it aims at maximising total (and therefore average) welfare: if one project generates more welfare than another, \rl{GreedCost} prefers that project. However, maximising total welfare does not take into account how `equal' of `fair' the budget allocation is, as the following example (adapted from \citet{peters2021proportional}) illustrates.

\begin{example}
Imagine a town consisting of four districts: North, East, South and West. All districts have roughly equal populations: North has 10.000 citizens, East has 9.900, South has 9.800 and West has 9.700 citizens. A budget of €10.000 is available for citizens' proposed projects. In every neighbourhood, many projects are submitted; in fact, the total overall budget is exceeded by each individual neighbourhood alone. Consequently, a voting round is organised.

It is not unnatural to assume that Northerners tend to vote mostly for projects in North, Easterners for projects in East, etc. Suppose that indeed, all Northerners vote for all and only all projects in North, all Easterners vote for those projects in East, etc. That means that all projects in North receive 10.000 votes, all projects in East receive 9.900 votes, all in South 9.800 votes and all in West 9.700 votes. The rule \rl{GreedCost} then selects a number of projects in North, after which it cannot fund any projects in East, South or West.
\end{example}

We see that in the above example, the entire budget allocation is decided by roughly a quarter of the voters, i.e., the budget allocation is in some sense `disproportional'. Multiple alternative PB rules have been proposed to achieve different notions of proportionality \citep{rey2023computational}. One such proposal by \citet{peters2021proportional} is the Method of Equal Shares, which we define in the following subsection.

\subsection{The Method of Equal Shares}

Although the Method of Equal Shares is defined for any satisfaction function, we assume again -- for simplicity and consistency with the above -- that a voter's welfare is proportional to the total cost of all selected projects that she approves of.

The intuition behind the Method of Equal Shares is the following. The total budget is equally divided among all voters (virtually) and we simulate the voters forming coalitions to `buy' a project that the whole coalition approves of, as follows. If there is any project that costs more than the budget share owned by all its approvers together, the project is disregarded. For all other projects, we find the project for which the cost can be `most equally distributed' over its approvers. This project is selected and the distributed cost is subtracted from each approver's virtual budget. We repeat this process until all projects are either selected or disregarded.

The term `most equally distributed' needs some specification. How equal a distribution is, in this case, is expressed by a parameter $\alpha\in\bR_{\geq 0}$ representing the ratio of the cost paid by and the welfare gained by the largest contributor to `buying' the project. That is, if a project has cost $c$ and has $k$ approvers, we aim to let each approver contribute $\frac{c}{k}$ to buying the project. If some approvers cannot afford this contribution, they contribute all of their remaining budget and the other approvers must contribute more than $\frac{c}{k}$, say $\gamma$, to compensate. The parameter $\alpha$ is then defined as $\frac{\gamma}{c}$, i.e., the ratio between the cost paid by the largest contributor (namely $\gamma$) and the welfare gained by the largest contributor if the project is selected (namely $c$). The lower $\alpha$, the more `equal' our distribution is.

Formally, the Method of Equal Shares is defined as follows.

\begin{definition}[Method of Equal Shares]
    The \emph{Method of Equal Shares} \rl{mes} (for cost satisfaction) is the PB rule generated by the following process.

    \begin{enumerate}
        \item Given a set of voters~$\cN = \{1,\dots,n\}$, a PB instance~$I = (\cP, c, b)$ and a profile $\tup{A}$, define for each agent $i\in\cN$ a virtual budget $b_i$. Initially, set each $b_i$ to $\frac{b}{n}$. Construct a subset $\pi$ of $\cP$ as follows. Initially, $\pi = \emptyset$.

        \item For each project $p\in\cP$, let the set of \emph{approvers} $\app{p}\subseteq\cN$ be the set of agents $i\in\cN$ such that $A_i(p) = 1$. If the project is not affordable by its approvers, i.e., $\sum_{i\in\app{p}}b_i < c(p)$, continue to the next project. Otherwise, initialise for each agent $i\in \app{p}$ her \emph{contribution to $p$} as $\gamma^p_i = \frac{c(p)}{|\app{p}|}$ (i.e., all approvers contribute equally to the cost of $p$).
        \begin{enumerate}
            \item We call agents $i\in\cN$ with $b_i > \gamma^p_i$ \emph{rich} agents. All other agents are \emph{poor} agents. Compute the sum of the poor agents' budgets ${s = \sum_{\{i\in\cN\mid b_i \leq \gamma^p_i\}}b_i}$ (this is the maximal contribution all poor agents can make together). For all poor agents $i$, update $\gamma^p_i$ to $b_i$. For all rich agents~$i$, update $\gamma^p_i$ to $\frac{c(p) - s}{|\{j\in\cN\mid b_j > \gamma^p_j\}|}$ (this is an equal distribution over the rich agents of the part of the project cost that cannot be covered by the poor agents).
            
            \item Repeat process (a) until $b_i\geq \gamma^p_i$ for all $i\in\app{p}$. The \emph{affordability of~$p$} is $\alpha^p := \frac{\max_{i\in\app{p}}\gamma^p_i}{c(p)}$, i.e., the affordability is the ratio between the maximal contribution made by an agent and the cost of the project.
        \end{enumerate}

        \item Select the project $p\in\cP$ with minimal affordability $\alpha^p$. Add $p$ to $\pi$ and update $b_i$ to $b_i - \gamma^p_i$ for each agent $i\in\app{p}$ (i.e., the approvers of $p$ collectively buy $p$ and distribute the cost as equally as possible).

        \item Repeat processes 2.\ and 3.\ until no project is affordable anymore. Return~$\pi$.
    \end{enumerate}
\end{definition}

Note that \rl{mes}, by design, respects at least some notion of proportionality: if a voter's virtual budget runs out (or gets smaller), she cannot contribute to projects anymore and therefore has no (or less) power to select projects she approves of. Therefore, a coalition of voters can never exert more decision power than their `fair' proportion of the budget. More precisely, \rl{mes} satisfies the axiom \emph{EJR up to one project} (EJR-1), which is (a relaxation of) a formalisation of proportionality \citep{peters2021proportional}.

\subsection{Completion of PB Rules}

The Method of Equal Shares has a drawback for most practical applications: it is not \emph{complete}. We call a PB rule complete if given any profile, the budget allocation $\pi$ it returns, cannot be extended by any project $p\in\cP\setminus\pi$ without exceeding the budget. The literature provides multiple approaches to `completing' a PB rule, two of which we consider in this project \citep{rey2023computational}. 

The first approach combines a given PB rule $R$ with a second rule $R'$ that is complete. Given a PB instance $I$ and a profile $\tup{A}$, we first apply the rule $R$ to $I$ and $\tup{A}$ to obtain a budget allocation $\pi$. Then we remove all projects $p\in\pi$ from $I$ and $\tup{A}$ and subtract the total cost of $\pi$ from the budget limit. We apply the rule $R'$ to the updated instance and profile to obtain a second budget allocation $\pi'$. Now, the budget allocation $\pi\cup\pi'$ is a feasible and complete budget allocation for the initial instance $I$. In the following, if we complete a PB rule~$R$ with \rl{GreedCost} as the secondary rule, we denote the resulting rule as $R^{+}$.

The second approach aims to complete a PB rule $R$ by iteratively running the rule $R$ on a PB instance $I$, where for each run, the budget of $I$ is increased by $\epsilon > 0$. If in some round $i\in\bN$, the resulting budget allocation $\pi_i$ is complete and feasible for the initial budget of $I$, we return $\pi_i$ and terminate. If the resulting budget allocation $\pi_i$ is not feasible for the initial budget of $I$, we return $\pi_{i-1}$ and terminate. Otherwise, we increase the budget by $\epsilon$ and continue to round~$i+1$. Note that this approach of completing a PB rule does not always yield a complete budget allocation. It is nonetheless of practical importance, since it avoids using a different PB rule and is therefore arguably more `in the spirit of' rule $R$ than the first approach. In the following, if we complete a PB rule $R$ by this approach, we denote the resulting rule as $R^{*}$. Naturally, $R^{{*}{+}}$ should be read as $(R^{*})^{+}$.

In particular, $\rl{mes}^{+}$ is a single run of \rl{mes} followed by a run of \rl{GreedCost} and $\rl{mes}^{{*}{+}}$ is an iterated run of \rl{mes} with increasing budgets followed by a run of \rl{GreedCost}. 

\section{Results}

Having defined the relevant notions of PB, we turn to the analysis of the data obtained from Amsterdam's PB instances. In this section, we first consider properties of the instances and profiles alone, such as the vote count, the project count and the project costs. We then turn to a comparison of the greedy cost welfare rule, \rl{GreedCost}, with two different completions of the Method of Equal Shares, $\rl{mes}^{+}$ and $\rl{mes}^{{*}{+}}$. We first consider some general properties, such as overlap in the outcome of the rules and the median cost of selected projects; and then consider some fairness properties, such as the average satisfaction score and the Gini coefficient of the satisfaction scores. We conclude by analysing some typical example instances more qualitatively.

The data received from the municipality of Amsterdam contained 43 PB instances, 4 of which lacked crucial data (either the ballots submitted or the cost of some projects). We disregarded these instances in our analysis. The remaining instances vary widely in the number of voters and the number of projects submitted. The smallest vote count is 66 and the largest 14411. The smallest project count is 3 and the largest is 97. As not to influence the statistical analysis disproportionally by outliers, we also disregarded all instances with fewer than 100 voters or fewer than 10 projects. The remaining 35 instances were used for our analysis. All the data used in this project, can be accessed via \url{http://pabulib.org/?city=Amsterdam}.

The code used to convert the data to the standard data format PaBuLib (see \citep{stolicki2020pabulib}), to run the different rules and to statistically analyse the results, can be found at \url{https://gitlab.com/pctnelissen/empirical-analysis-of-participatory-budgeting}.

\label{sec:results}
\subsection{Properties of the Instances and Profiles}

\begin{figure}[h]
    \centering
    \includegraphics[width=0.9\textwidth]{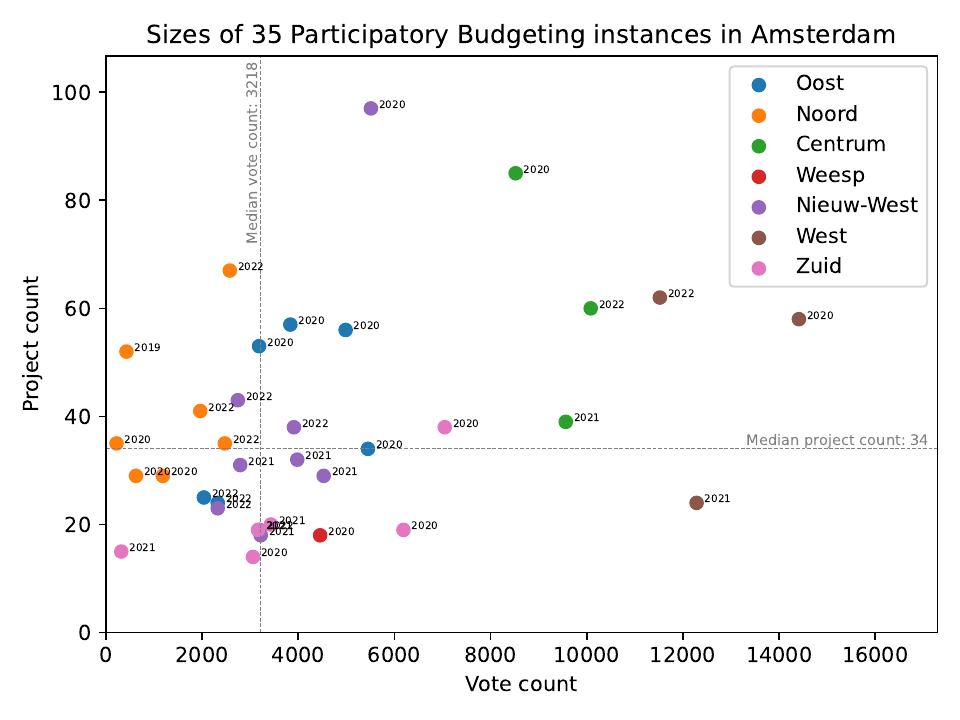}
    \caption{Project and vote counts for 35 Participatory Budgeting instances organised in Amsterdam between 2019 and 2022.}
    \label{fig:scatter}
\end{figure}

In Figure~\ref{fig:scatter}, the vote and project counts of all instances are plotted against each other. The median vote count is 3218 and the median project count is 34.\footnote{Both medians are plotted as a line, partitioning the instances in categories `small' and `large' per axis. All results in this section were studied for the data set as a whole, as well as partitioned into these size-based quadrants. However, all findings for the four quadrants were analogous to the findings for the data set as a whole, and are therefore omitted from this text.} All instances are coloured by the district in which they were organised and labelled by year. Note that some districts organised multiple PB instances in a single year. In those cases, the district was partitioned into multiple neighbourhoods and one PB instance was organised per neighbourhood.

\begin{figure}[b!]
    \centering
    \begin{subfigure}{0.49\textwidth}
        \centering
        \includegraphics[width=\textwidth]{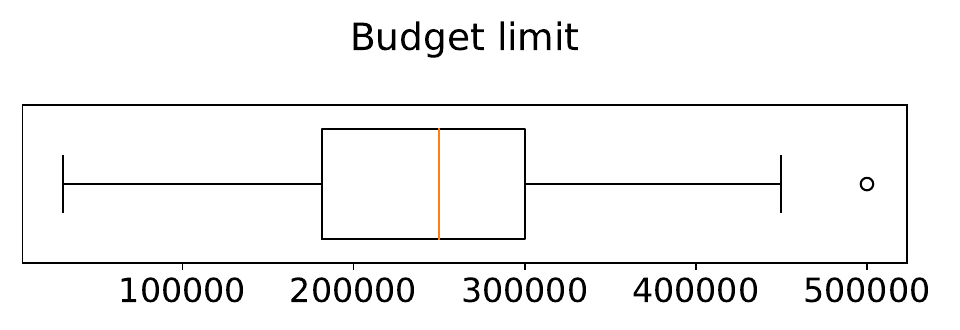}
        \caption{}
        \label{fig:budget}
    \end{subfigure}
    \hfill
    \begin{subfigure}{0.49\textwidth}
        \centering
        \includegraphics[width=\textwidth]{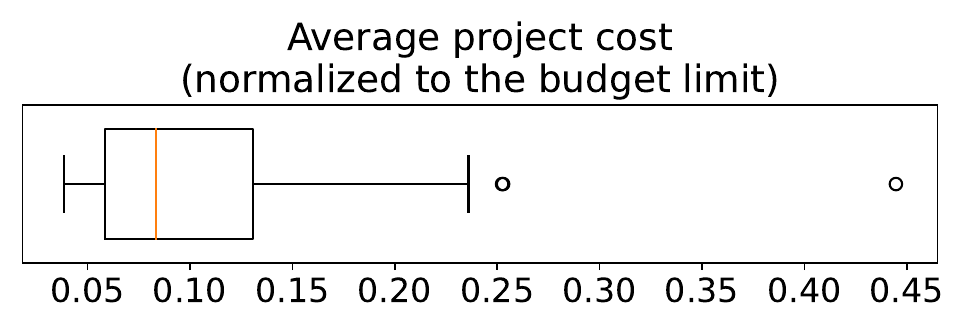}
        \caption{}
        \label{fig:project_cost}
    \end{subfigure}
    
    \vspace{12pt}
    \begin{subfigure}{0.49\textwidth}
        \centering
        \includegraphics[width=\textwidth]{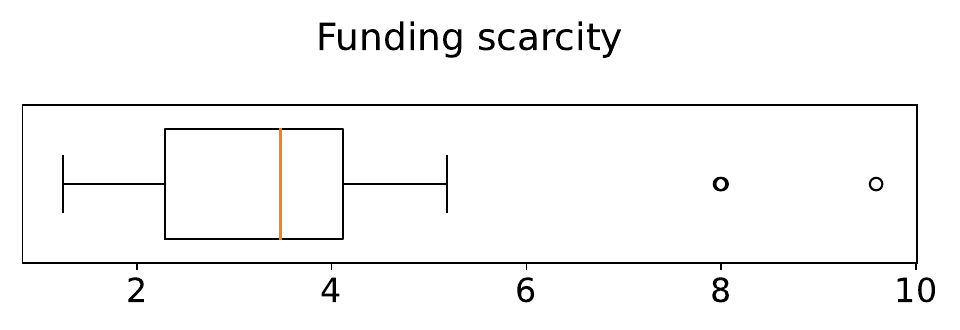}
        \caption{}
        \label{fig:scarcity}
    \end{subfigure}
    \hfill
    \begin{subfigure}{0.49\textwidth}
        \centering
        \includegraphics[width=\textwidth]{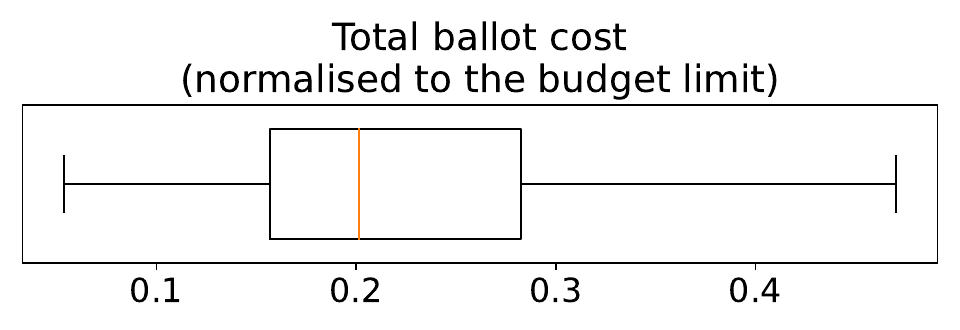}
        \caption{}
        \label{fig:ballot_cost}
    \end{subfigure}
    \caption{Four boxplots on instance and profile properties of 35 Participatory Budgeting instances.}
    \label{fig:boxplots}
\end{figure}

Figure~\ref{fig:boxplots} displays some properties of the budget limits, projects and ballots of the 35 PB instances. In Figure~\ref{fig:budget}, we see that the budget limits range from €30.000 to €500.000 with a median of €250.000. In Figure~\ref{fig:project_cost}, we see the average project cost per instance, normalised by the budget limit of the instance. The average project cost ranges from 4\% of the budget limit to 44\%, with a median of 8\%. In Figure~\ref{fig:scarcity}, we see a measure of funding scarcity for each instance, namely the ratio between the total sum cost of all projects in an instance and the budget limit in the instance. The instance with least funding scarcity has a budget limit of 1.2 times the total funding requested by all projects, and the highest funding scarcity is 9.6. The median is 3.5. Finally in Figure~\ref{fig:ballot_cost}, we see the average size of the ballots submitted in an instance, measured by the total cost of its approved projects and normalised by the budget limit of the instance. In the instance with the smallest average ballot, the average ballot approved projects costing 5\% of the budget limit, whereas the average ballot approved 47\% in the instance with the largest average ballot. The median is 20\%. Note that for many instances, voters were bound by some cardinality or cost constraints while voting, which clearly influences the latter boxplot.

\newpage\subsection{Properties of the Greedy Rule and the Method of Equal Shares}

We turn our attention to the PB rules \rl{GreedCost}, $\rl{mes}^{+}$ and $\rl{mes}^{{*}{+}}$, and analyse their performance on our data set.

\subsubsection{General Properties}
\label{sec:general}

\begin{figure}[b!]
    \centering
    \includegraphics[width=0.8\textwidth]{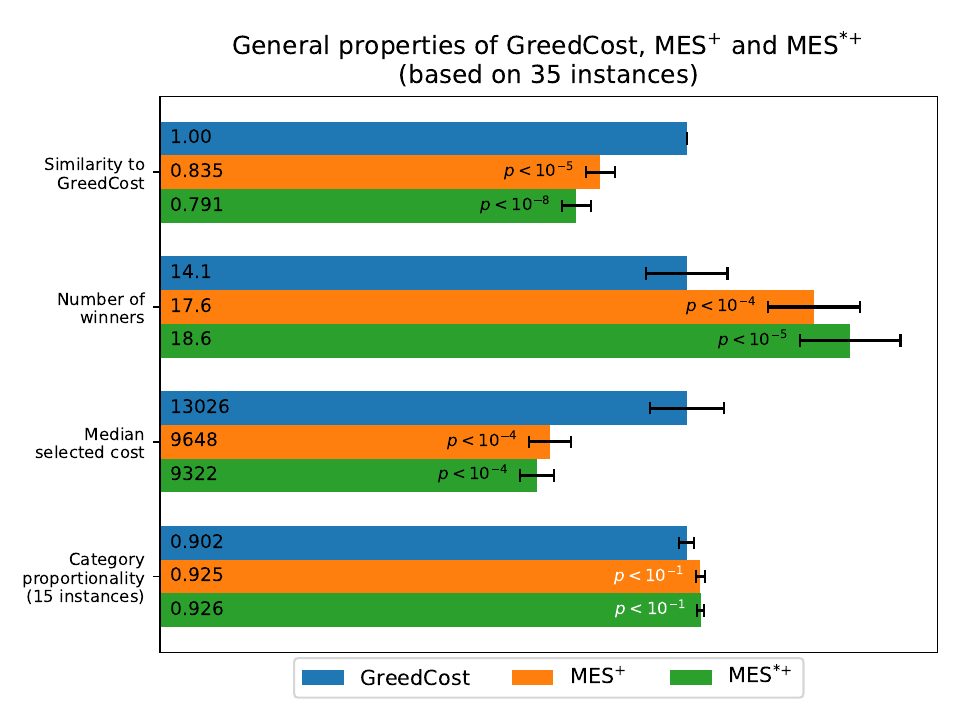}
    \caption{Four general properties of the PB rules \rl{GreedCost}, $\rl{mes}^{+}$ and $\rl{mes}^{{*}{+}}$. The error bars show the standard error. The $p$-values are obtained by a paired t-test between \rl{GreedCost} and the rule in question. We consider $p < 0.05$ statistically significant.}
    \label{fig:general}
\end{figure}

Figure~\ref{fig:general} plots four properties of the three PB rules. The \emph{similarity} between two different sets of winners $W_1$ and $W_2$ for the same PB instance is defined as \[\frac{\cost(W_1\cap W_2)}{\frac{1}{2}(\cost(W_1) + \cost(W_2))},\] where $\cost(W)$ is the total cost of all projects in a set $W$. Note that if $W_1$ and $W_2$ are complete (in the sense that adding another project to $W_i$ exceeds the budget limit), then $\frac{1}{2}(\cost(W_1) + \cost(W_2))$ is close to the budget limit. Further note that the similarity of $W_1$ and $W_2$ is 1 if and only if they are the same set, and it is 0 if and only if the sets are disjoint. We empirically measure the similarity of a rule $R$ to \rl{GreedCost} by the average similarity over all PB instances between the winners selected by $R$ and the winners selected by \rl{GreedCost}. Thus, in Figure~\ref{fig:general} we read that $\rl{mes}^{+}$ allocates an average of 17\% of the budget differently than \rl{GreedCost} does, and for $\rl{mes}^{{*}{+}}$ this is 21\%.

Furthermore, we see that the number of winners selected by $\rl{mes}^{+}$ and $\rl{mes}^{{*}{+}}$ is higher than for \rl{GreedCost}; and, connected to this, the median cost of the projects selected by $\rl{mes}^{+}$ or $\rl{mes}^{{*}{+}}$ is lower than the median cost selected by \rl{GreedCost}.

Finally, 15 of the PB instances in our data set contain categorisations of the submitted projects; for example, one instance differentiates between projects aimed at youth, projects aimed at green spaces and projects aimed at social cohesion. For these instances, we can compare the distribution of funding for projects in the different categories found in the ballots, versus the distribution of funding by a PB rule $R$. Given a profile $\tup{A}$ and a set of projects $C\subseteq\cP$, the proportion of funds allocated to $C$ by the voters is \[q(C) := \frac{1}{n}\sum_{i\in\cN}\frac{\cost(C\cap A_i)}{\cost(A_i)}.\] Given a PB rule $R$, the proportion of funds allocated to $C$ by $R$ is \[q_R(C) := \frac{\cost(C\cap R(\tup{A}))}{\cost(R(\tup{A}))}.\] We empirically measure the \emph{category disproportionality} of a rule $R$ by taking the square root of the mean squared difference between $q(C)$ and $q_R(C)$ for all categories $C\subseteq\cP$ identified in an instance. The \emph{category proportionality}, being the inverse of category disproportionality, is obtained by applying $x\mapsto e^{-x}$. Note that in Figure~\ref{fig:general}, category proportionality is the only measure that does not significantly change from \rl{GreedCost} to $\rl{mes}^{+}$ or $\rl{mes}^{{*}{+}}$.

\subsubsection{Fairness Properties}
\label{sec:fairness}

\begin{figure}[h]
    \centering
    \includegraphics[width=0.8\textwidth]{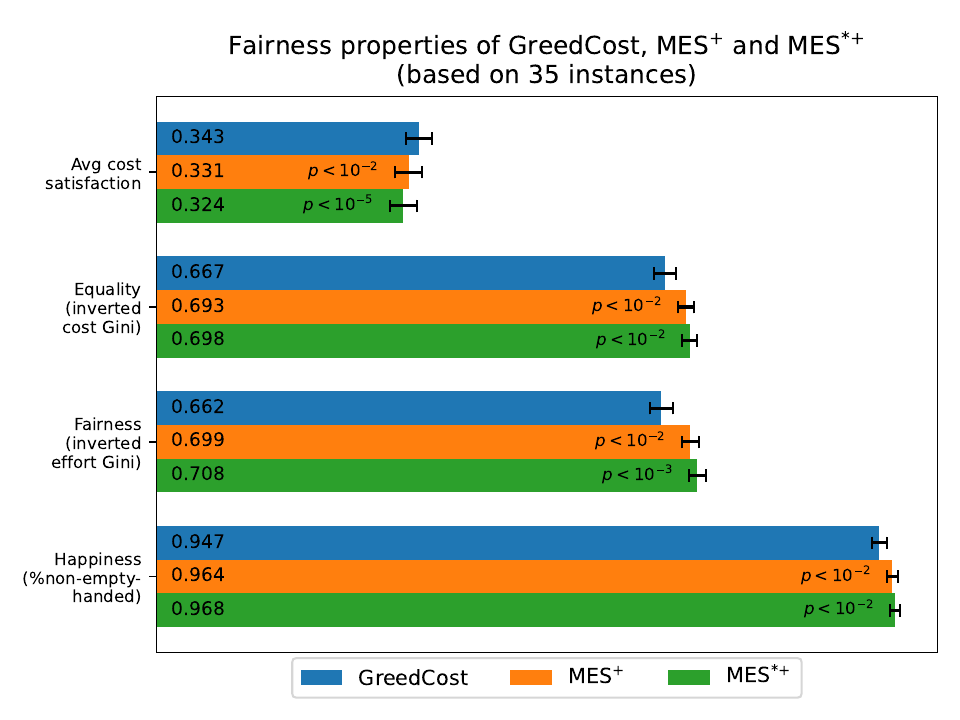}
    \caption{Four fairness properties of the PB rules \rl{GreedCost}, $\rl{mes}^{+}$ and $\rl{mes}^{{*}{+}}$. The error bars show the standard error. The $p$-values are obtained by a paired t-test between \rl{GreedCost} and the rule in question. We consider $p < 0.05$ statistically significant.}
    \label{fig:fairness}
\end{figure}

PB rules can be compared by different measures of fairness. Figure~\ref{fig:fairness} displays four such measures. Given a budget limit $b$, a profile $\tup{A}$ and a PB rule $R$, the average cost satisfaction is the average of $\frac{\cost(R(\tup{A})\cap A_i)}{b}$ over all voters $i\in\cN$. That is, the cost satisfaction of a voter is the total cost of the selected projects that the voter approves of, normalised by the budget limit. Figure~\ref{fig:fairness} plots the mean average cost satisfaction over all instances in our data set.

Furthermore, Figure~\ref{fig:fairness} plots the Gini coefficient (a well-known measure of the equality of a distribution of resources) of the cost satisfaction of all voters, and the Gini coefficient of the \emph{effort} (or \emph{share}) welfare measure. The latter is a notion introduced by \citet{lackner2021fairness} to measure the `effort' put in by decision makers to satisfy a voter. As such, the effort Gini coefficient expresses to what extent the decision makers distribute their `efforts' fairly.

Finally, we call a voter \emph{happy} if at least one of the projects she approves of, is selected. The overall happiness of voters is the percentage of voters that are happy, i.e., that are not left empty-handed by the PB rule.

\subsection{Qualitative Examples}

To illustrate qualitatively the differences between the three rules considered above, we study three example instances. The examples were selected to illustrate the largest negative effect that switching from \rl{GreedCost} to $\rl{mes}^{{*}{+}}$ would have on fairness and category proportionality, the median effect it would have and the largest positive effect it would have (within the instances of the data set that contain category information). The examples were selected by taking the minimal, median and maximal average increase of equality and category proportionality, as defined in Sections~\ref{sec:general} and \ref{sec:fairness}.

\subsubsection*{Instance 522: The Largest Negative Effect}

\begin{figure}[b!]
   \begin{subfigure}[T]{0.49\textwidth}
        \includegraphics[width=\textwidth]{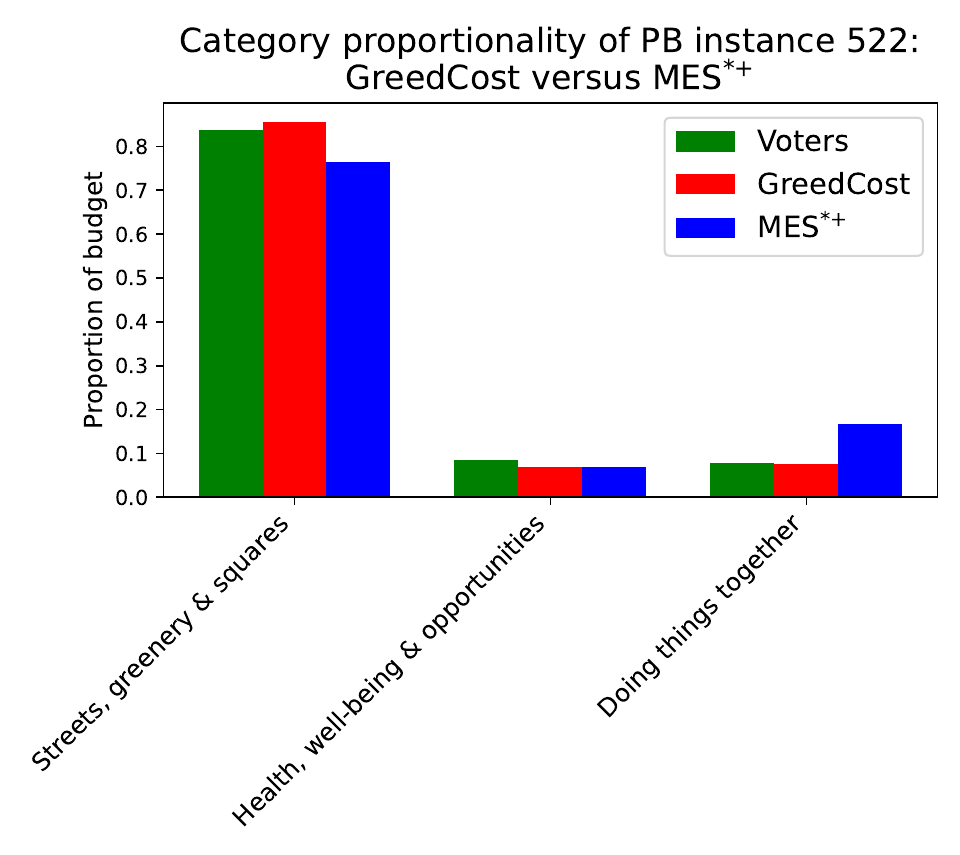}
        \caption{}
        \label{fig:522_prop}
    \end{subfigure}
    \hfill
    \begin{subfigure}[T]{0.49\textwidth}
        \includegraphics[width=\textwidth]{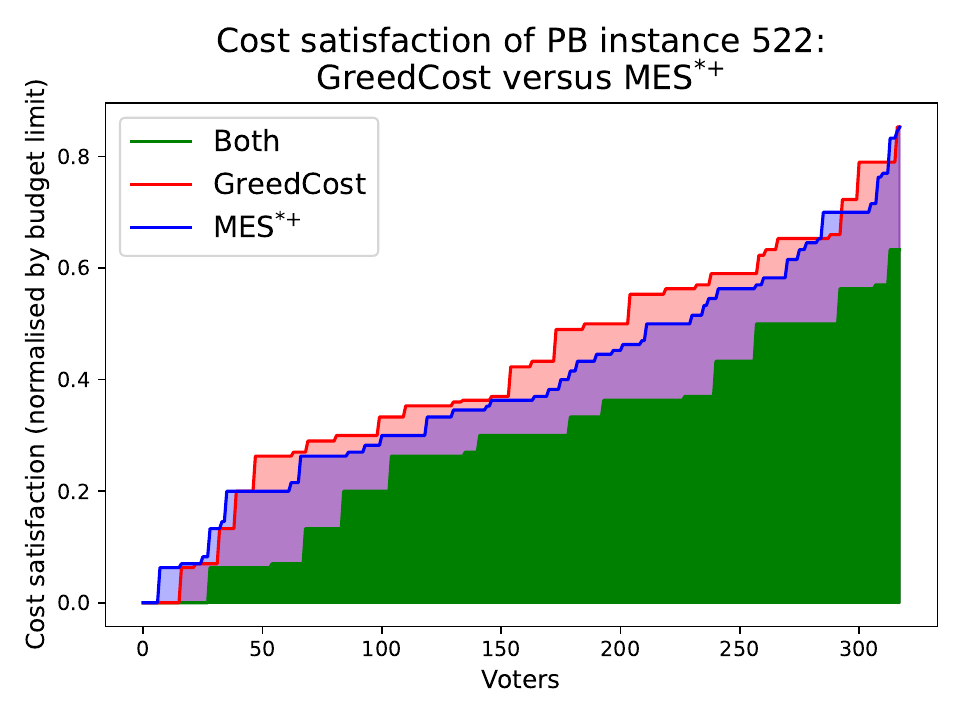}
        \caption{}
        \label{fig:522_cost}
    \end{subfigure}
    
    \caption{Category proportionality and cost satisfaction distribution of instance 522. This is the instance with the largest negative effect when switching from \rl{GreedCost} to $\rl{mes}^{{*}{+}}$.}
    \label{fig:522}
\end{figure}

The largest negative effect of switching from \rl{GreedCost} to $\rl{mes}^{{*}{+}}$ occurred in the instance in neighbourhood Diamantbuurt in 2021, identified by ID 522. The following list summarises how many projects were selected in each category by the different rules.\footnote{The exact projects selected by the rules in instance 522 can be found in Appendix~\ref{app:522}.}

\begin{itemize}
\item Selected by both rules (€63.300):
\begin{itemize}
\item Streets, greenery \& squares: 2 projects (€50.000)
\item Doing things together: 1 project (€7.000)
\item Health, well-being \& opportunities: 1 project (€6.300)
\end{itemize}
\item Selected by \rl{GreedCost} only (€29.000):
\begin{itemize}
\item Streets, greenery \& squares: 1 project (€29.000)
\end{itemize}
\item Selected by $\rl{mes}^{{*}{+}}$ only (€28.250):
\begin{itemize}
\item Streets, greenery \& squares: 1 project (€20.000)
\item Doing things together: 1 project (€8.250)
\end{itemize}
\end{itemize}

Note that the largest part of the budget, €63.300, is allocated identically by \rl{GreedCost} and $\rl{mes}^{{*}{+}}$. The rule \rl{GreedCost} supplements this common allocation by one project of category `Streets, greenery \& squares', while $\rl{mes}^{{*}{+}}$ supplements the common allocation by two cheaper projects from the categories `Streets, greenery \& squares' and `Doing things together'.

Figure~\ref{fig:522_prop} demonstrates the category proportionality of \rl{GreedCost} and $\rl{mes}^{{*}{+}}$. In green, we see the distribution of funds over the different categories directly by the voters, and in red and blue we see the distribution that the two PB rules bring about. The closer the red or blue values are to the green values, the more proportional the allocation is. In the case of PB instance 522, we see that \rl{GreedCost} (surprisingly) performs slightly better than $\rl{mes}^{{*}{+}}$.

Figure~\ref{fig:522_cost} demonstrates the distribution of the cost satisfaction over all voters: the $x$-axis enumerates all voters from lowest cost satisfaction to highest cost satisfaction, and the value plotted is the total cost of all projects selected by the respective rule that the voter approves of. In green, we see this distribution if only those projects are funded that are selected by both \rl{GreedCost} and $\rl{mes}^{{*}{+}}$. As such, the red and blue areas illustrate the unique contribution that the respective rule makes to the satisfaction distribution. Note that for voters with low satisfaction, $\rl{mes}^{{*}{+}}$ is preferable to \rl{GreedCost}, whereas for most other voters it is the other way around.

\subsubsection*{Instance 613: The Median Effect}

The median effect of switching from \rl{GreedCost} to $\rl{mes}^{{*}{+}}$ occurred in the instance in neighbourhood Noord Oost in 2022, identified by ID 613. We analyse the example analogously to the above. The projects selected in the different categories are the following.\footnote{The exact projects selected by the rules in instance 613 can be found in Appendix~\ref{app:613}.}

\begin{itemize}
\item Selected by both rules (€201.678):
\begin{itemize}
\item Equal opportunities \& education: 4 projects (€53.765)
\item Meeting \& connection: 4 projects (€49.130)
\item Youth activities: 2 projects (€46.065)
\item Sports \& exercise: 2 projects (€28.938)
\item Safety: 2 projects (€23.780)
\end{itemize}
\item Selected by \rl{GreedCost} only (€59.000):
\begin{itemize}
\item Meeting \& connection: 1 project (€59.000)
\end{itemize}
\item Selected by $\rl{mes}^{{*}{+}}$ only (€57.781):
\begin{itemize}
\item Equal opportunities \& education: 1 project (€5.404)
\item Meeting \& connection: 4 projects (€19.782)
\item Sports \& exercise: 5 projects (€26.095)
\item Art \& creativity: 1 project (€6.500)
\end{itemize}
\end{itemize}

Note that in this case, $\rl{mes}^{{*}{+}}$ replaces a single project selected by \rl{GreedCost} by a list of 11 projects in diverse categories.

\begin{figure}[t]
   \begin{subfigure}[T]{0.49\textwidth}
        \includegraphics[width=\textwidth]{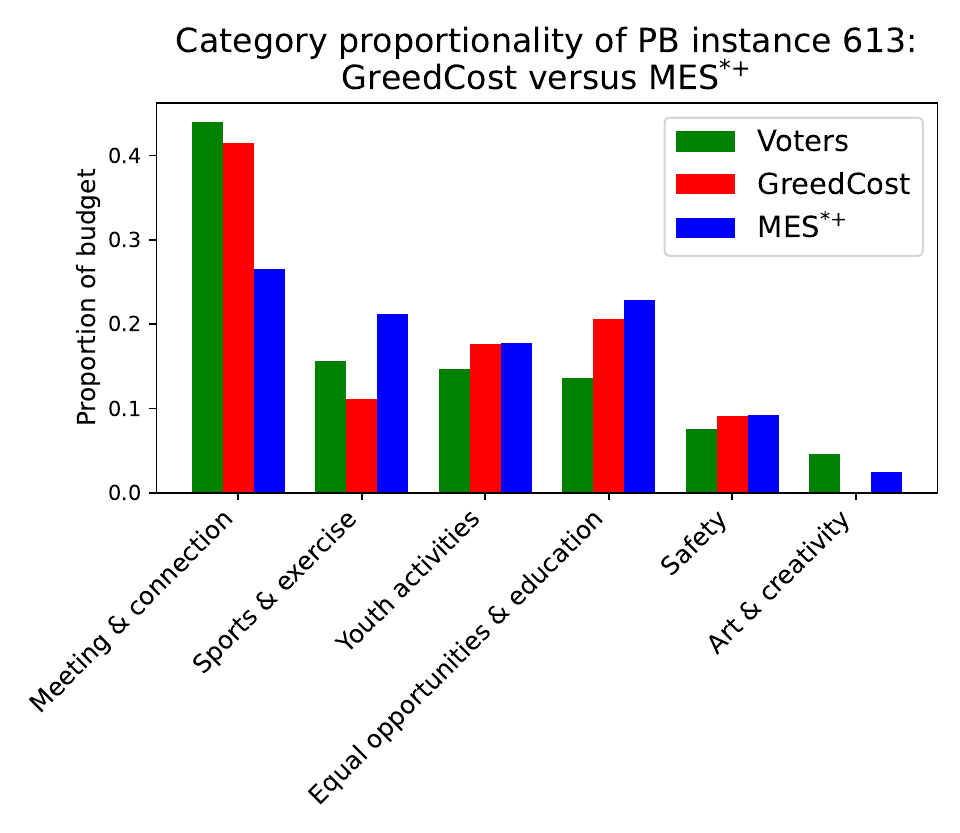}
        \caption{}
        \label{fig:613_prop}
    \end{subfigure}
    \hfill
    \begin{subfigure}[T]{0.49\textwidth}
        \includegraphics[width=\textwidth]{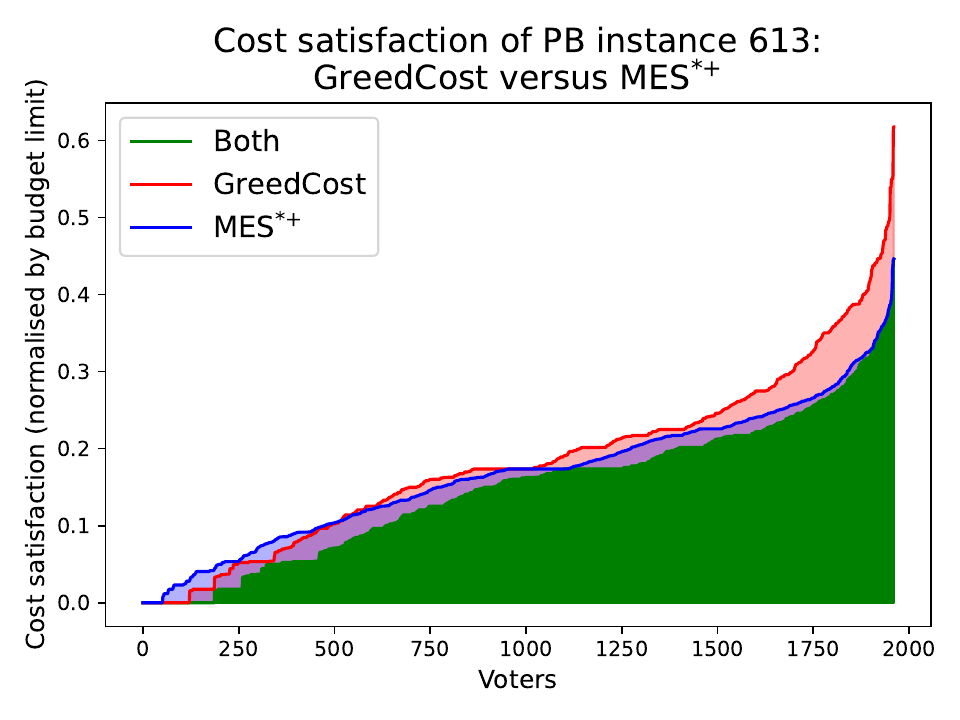}
        \caption{}
        \label{fig:613_cost}
    \end{subfigure}
    
    \caption{Category proportionality and cost satisfaction distribution of instance 613. This is the instance with the median effect when switching from \rl{GreedCost} to $\rl{mes}^{{*}{+}}$.}
    \label{fig:613}
\end{figure}

In Figure~\ref{fig:613_prop}, we see that $\rl{mes}^{{*}{+}}$ is slightly more proportional than \rl{GreedCost} for instance 613. And in Figure~\ref{fig:613_cost}, we see again that voters with low satisfaction should prefer $\rl{mes}^{{*}{+}}$, whereas most voters should prefer \rl{GreedCost}.

\newpage\subsubsection*{Instance 644: The Largest Positive Effect}

Finally, the largest positive effect of switching from \rl{GreedCost} to $\rl{mes}^{{*}{+}}$ occurred in the instance in neighbourhood Geuzenveld Slotermeer in 2022, identified by ID 644. The projects selected in the different categories are the following.\footnote{The exact projects selected by the rules in instance 644 can be found in Appendix~\ref{app:644}.}

\begin{itemize}
\item Selected by both rules (€118.500):
\begin{itemize}
\item Public space: 3 projects (€101.500)
\item Social: 1 project (€7.000)
\item Green: 1 project (€5.000)
\item Other: 1 project (€5.000)
\end{itemize}
\item Selected by \rl{GreedCost} only (€147.000):
\begin{itemize}
\item Public space: 2 projects (€147.000)
\end{itemize}
\item Selected by $\rl{mes}^{{*}{+}}$ only (€147.260):
\begin{itemize}
\item Public space: 2 projects (€10.000)
\item Social: 7 projects (€92.760)
\item Green: 2 projects (€17.000)
\item Culture: 2 projects (€21.500)
\item Other: 1 project (€6.000)
\end{itemize}
\end{itemize}

Note that in this case, $\rl{mes}^{{*}{+}}$ replaces two projects of the same category selected by \rl{GreedCost} (representing more than half of the budget) by a list of 14 projects in diverse categories.

\begin{figure}[h]
   \begin{subfigure}[T]{0.49\textwidth}
        \includegraphics[width=\textwidth]{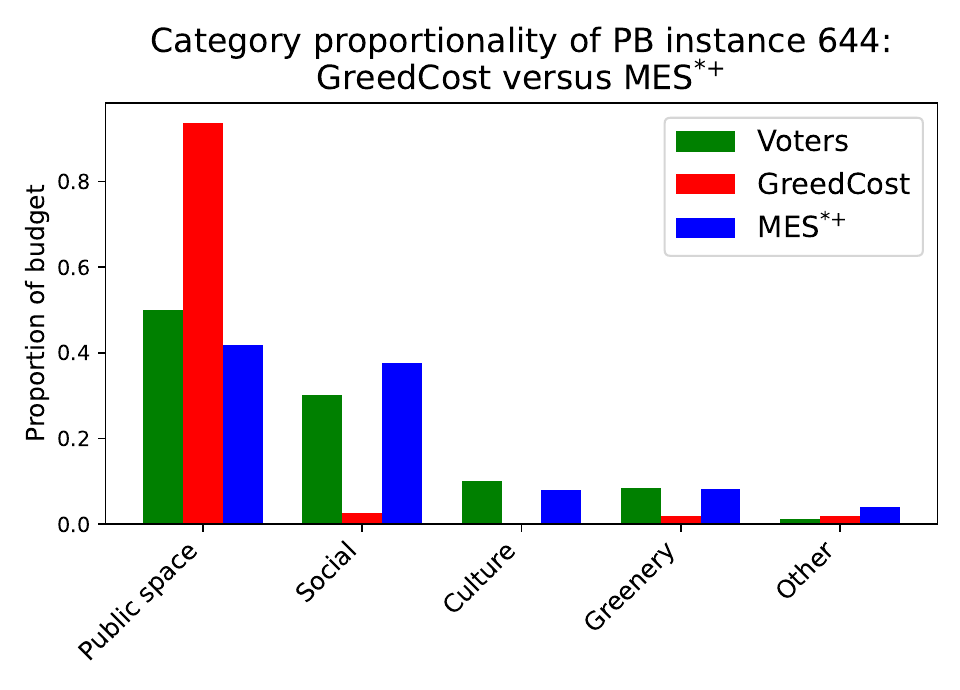}
        \caption{}
        \label{fig:644_prop}
    \end{subfigure}
    \hfill
    \begin{subfigure}[T]{0.49\textwidth}
        \includegraphics[width=\textwidth]{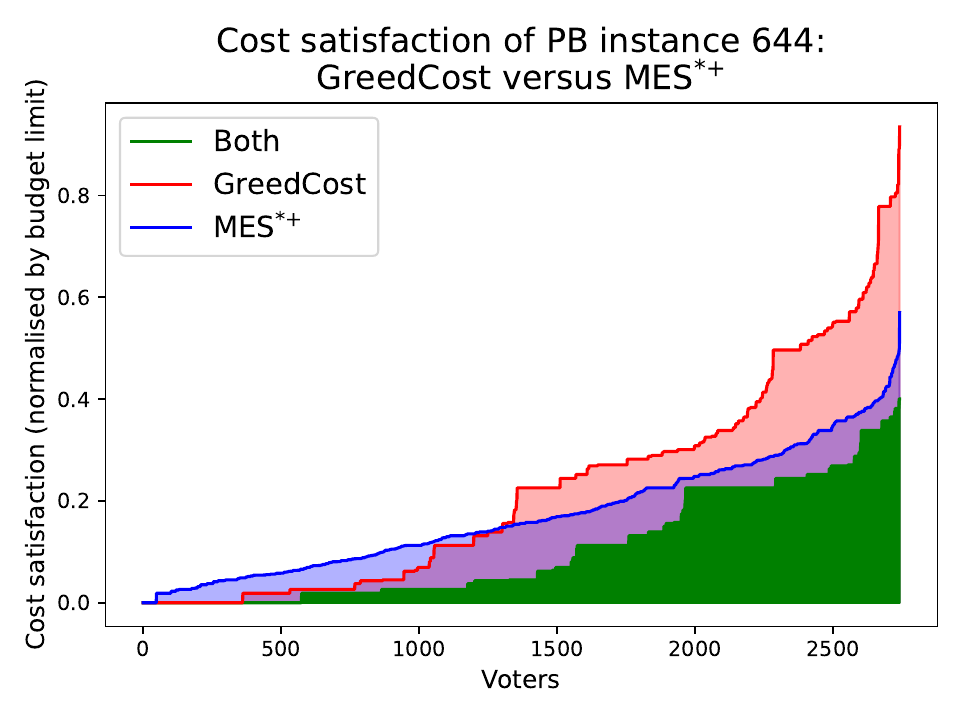}
        \caption{}
        \label{fig:644_cost}
    \end{subfigure}
    
    \caption{Category proportionality and cost satisfaction distribution of instance 644. This is the instance with the largest positive effect when switching from \rl{GreedCost} to $\rl{mes}^{{*}{+}}$.}
    \label{fig:644}
\end{figure}

In Figure~\ref{fig:644_prop}, we see that $\rl{mes}^{{*}{+}}$ is far more proportional than \rl{GreedCost} for instance 644. And in Figure~\ref{fig:644_cost}, we see a similar effect as above, but stronger: voters with low satisfaction should prefer $\rl{mes}^{{*}{+}}$, whereas voters with high satisfaction should prefer \rl{GreedCost}.

\section{Conclusion and Discussion}
\label{sec:conclusion}
We analysed the greedy cost welfare and two different completions of the Method of Equal Shares, using 35 PB instances conducted in Amsterdam between 2019 and 2022. Our primary conclusion is that the difference in budget allocation by these rules is significant: $\rl{mes}^{+}$ allocates 17\% of the budget differently from \rl{GreedCost}, and for $\rl{mes}^{{*}{+}}$ this is 21\%. Which rule should be preferred, depends on the aims of a particular PB instance. However, the results obtained in this project can offer guidance to authorities organising a PB instance.

In terms of fairness, Figure~\ref{fig:fairness} shows that the Method of Equal Shares generates results that can be characterised as `more fair' or `more equal' than the greedy cost welfare rule, at the expense of total (i.e., average) voter satisfaction. The slightly increased average equality and fairness of the outcome under the Method of Equal Shares, is due to an increase of (cost and effort) satisfaction for a minority at the bottom of the distribution, at the expense of the satisfaction of the majority at the top of the distribution. The increase of happiness from 94\% to 96\% or 97\% seems a rather small increase at first sight, but it roughly halves the number of citizens that do not approve of a single selected project. For the median vote count of roughly 3.000, this translates to about 70 extra voters not being left empty-handed. Arguably, the increase of three different measures of fairness outweighs the slightly decreased average cost satisfaction.

Other properties of the PB rules might also be relevant for designing a PB process: category proportionality, median selected cost and explainability. The category proportionality of the Method of Equal Shares seems to be higher than that of the greedy cost welfare rule, but the difference is not statistically significant in our data set. However, from the examples studied above, it seems that the Method of Equal Shares selects a more diverse set of projects than the greedy cost welfare rule, resulting in funding a set of projects that is more representative of the voters' preferences. As a result of selecting more (and thus smaller) projects, the median selected cost by the Method of Equal Shares is considerably lower than by the greedy cost welfare rule. Whether benefiting cheaper projects over more expensive projects is an asset or a drawback, is highly context-dependant. Finally, the Method of Equal Shares is harder to explain and understand than the greedy cost welfare rule. As a consequence, voters might have more trust in the greedy cost welfare rule, because they better understand how the budget allocation is decided. For this reason, a popular explanation of the Method of Equal Shares is included in the appendix.

All in all, we conclude that the Method of Equal Shares and the greedy cost welfare rule yield considerably different budget allocations. The Method of Equal Shares arguably allocates the budget more fairly and equally, and selects a more proportional, diverse set of cheaper projects; all at the expense of a slight decrease in average voter satisfaction.

\bibliography{bib.bib}

\newpage\appendix
\section{A Popular Explanation of the Method of Equal Shares (English)}

\subsection{Visual Explanation}

In Participatory Budgeting (PB), we aim to democratically distribute a budget over some items. For example, six housemates have \texteuro 90 to spend on a vacuum cleaner, a casserole, a mixer and a plant, but they cannot afford all items. Each housemate submits a ballot stating which items they would like. A \emph{Participatory Budgeting rule} decides which items are funded.

\begin{figure}[h!]
\centering
\vspace{12pt}
\includegraphics[width=0.61\textwidth]{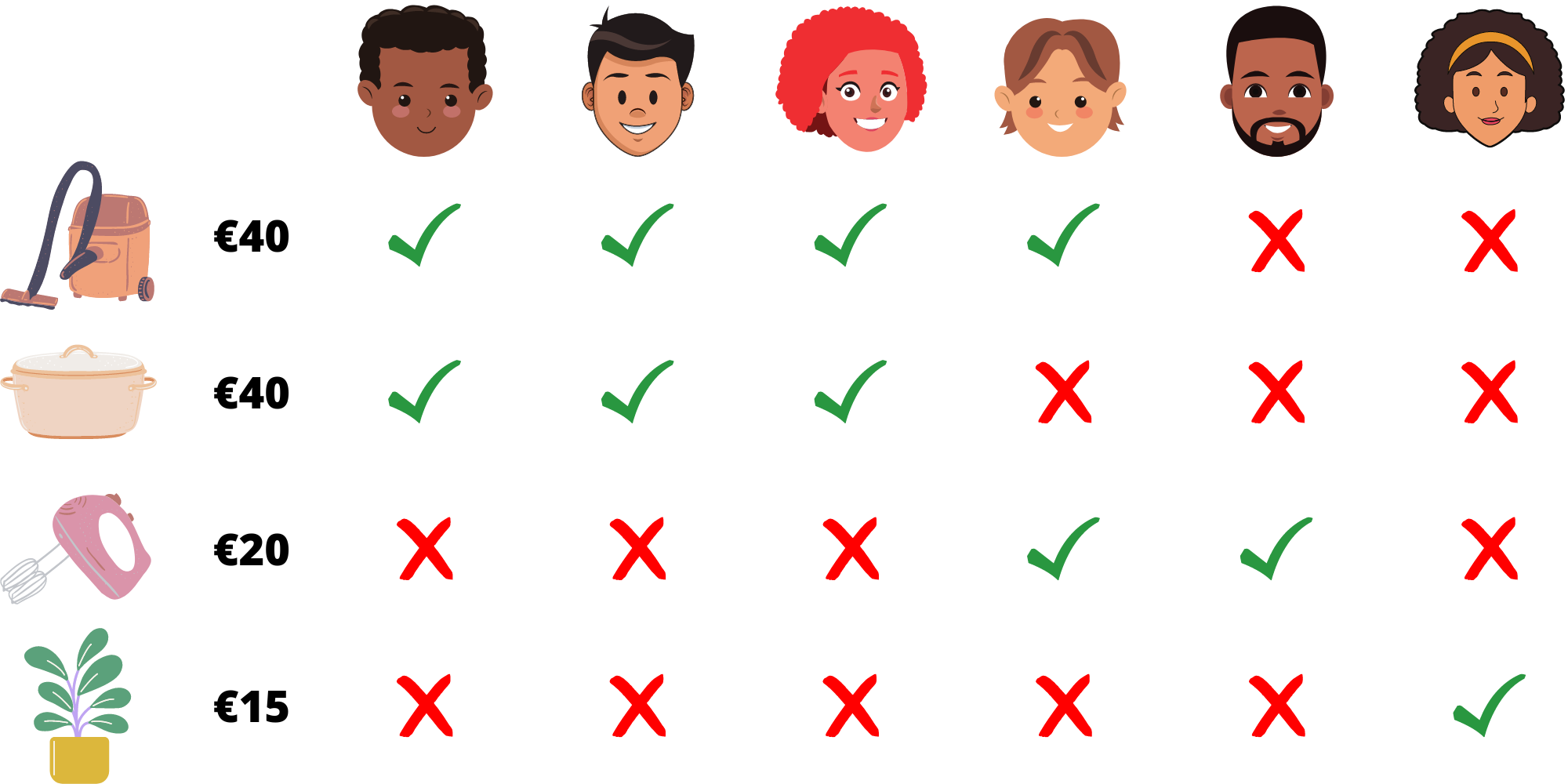}
\end{figure}

The \emph{Greedy Cost Welfare} (\rl{GreedCost}) selects the items with the most votes: the vacuum and the casserole.

\begin{figure}[h!]
\centering
\vspace{15pt}
\includegraphics[width=0.61\textwidth]{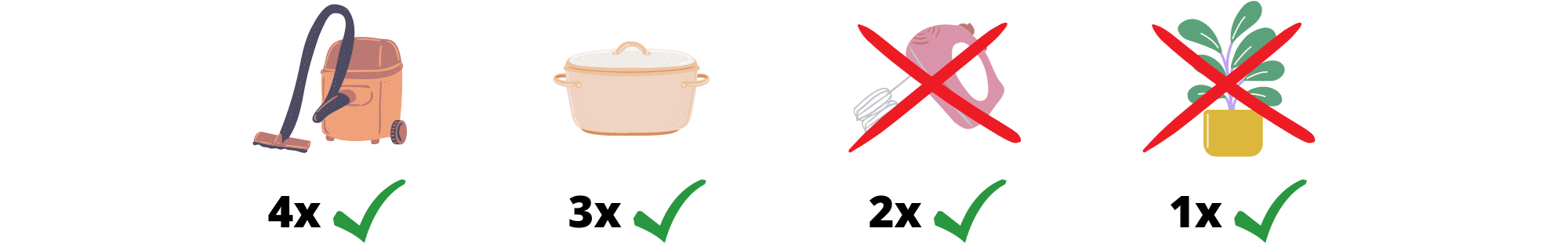}
\end{figure}

The \emph{Method of Equal Shares} (\rl{MES}) distributes the budget evenly over all voters, and simulates coalitions buying the items for which the cost can be distributed most evenly. It selects the vacuum, mixer and plant. Note that now no voter is left empty-handed.

\begin{figure}[h!]
\centering
\vspace{12pt}
\includegraphics[width=0.61\textwidth]{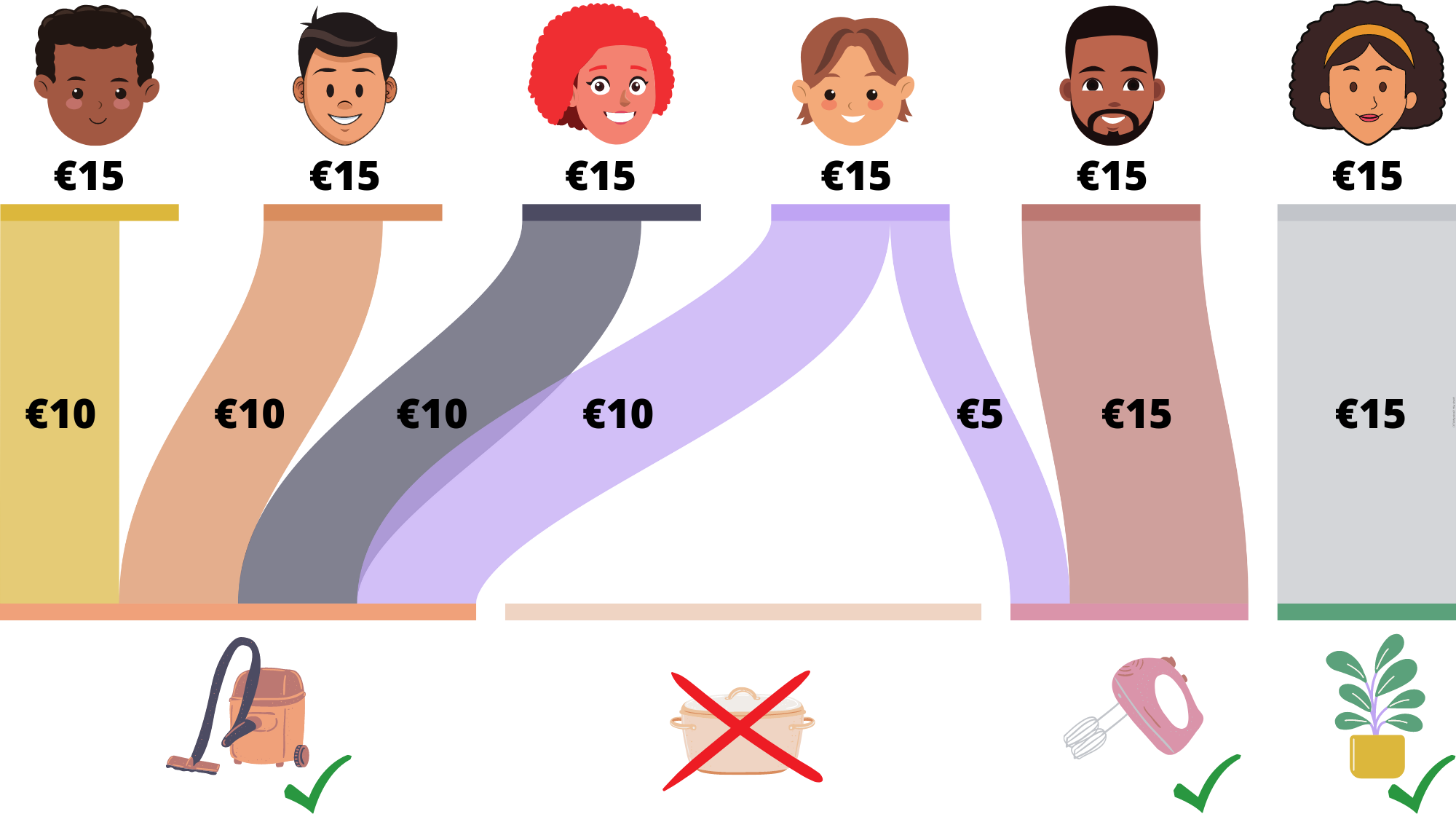}
\end{figure}

\subsection{Textual Explanation}

The goal of Participatory Budgeting (PB) is to democratically decide which projects a government should fund and which projects it should reject. A PB process consists of collecting possible projects, organising an election, and announcing which projects will be funded. Between the last two steps, votes are counted and winning projects are selected. This can be done in multiple ways, including the `Method of Equal Shares' (MES). Below, we explain how MES works. Another explanation can be found at \url{https://equalshares.net/explanation}.

The aim of MES is to distribute the budget as fairly as possible; every voter should have an equal influence on the outcome. The voting rule tries to accomplish this by virtually distributing the budget among all voters, and allowing voters to `buy' projects together. When a voter's virtual wallet is empty, the voter no longer has any influence on which additional projects are funded.

Voting is done with approval ballots. This means that each voter indicates which projects they approve of and which projects they reject. The MES rule is then used to count the votes.

MES can be seen as a simulation of voters forming coalitions to `buy' projects together. The simulation consists of the following steps.
\begin{enumerate}
\item The total budget is evenly distributed among all voters: each voter has a virtual wallet with an equal amount of money.
\item For each project, we compute whether the approvers of the project have enough money in their virtual wallets to buy the project together. If not, the project is not funded.
\item If multiple projects remain, we consider how \emph{fairly} the costs of each project can be distributed among its approvers. The project for which the costs can be distributed most fairly, is bought by its approvers. The costs are subtracted from the virtual wallets, and the simulation returns to step 2.
\item When each project has been rejected or bought, the simulation ends.
\end{enumerate}

How \emph{fairly} the costs of a project can be distributed, is determined as follows.

Initially, the costs of a project are distributed equally among its approvers. However, some voters might have already `bought' other projects and, therefore, might not have enough money left to pay this equal share. If so, these voters pay their entire remaining budgets. The costs they cannot cover, are divided equally among the other approvers that still have enough money in their wallets.

As such, the distribution of the costs becomes somewhat unequal. The inequality of the distribution is calculated by dividing the contribution by each voter, $\gamma$, by the total cost of the project, $c$. We thus obtain a value ${\alpha = \frac{\gamma}{c}}$ for each voter. Assuming that the total cost of a project is equal to the satisfaction the project provides, $\alpha$ is the cost per unit satisfaction paid by a voter. When the costs can be distributed perfectly equally, $\alpha$ is low for all approvers of a project. When the costs are distributed very unequally, the $\alpha$ of the largest contributor is much larger than that of smaller contributors. The cost distribution of the project with the lowest maximum $\alpha$ is therefore considered the most \emph{fair}.

All in all, MES simulates what voters would do if they were all given an equal share of the budget, and were allowed to (as fairly as possible) form coalitions to realise projects. In this way, each voter has an equal influence on the selection of projects, and we achieve a proportional and fair budget allocation.

\newpage\section{A Popular Explanation of the Method of Equal Shares (Dutch)}

\subsection{Visuele uitleg}

Het doel van Participatieve Begroting (PB) is om op democratische wijze een begroting op te stellen. Bijvoorbeeld, zes huisgenoten hebben €90 te besteden aan een stofzuiger, een pan, een mixer en een plant, maar ze kunnen niet alle items betalen. Dus geeft elke huisgenoot aan welke items ze willen hebben. Een \emph{Participatieve Begrotingsregel} beslist welke items ze kopen.

\begin{figure}[h!]
\centering
\vspace{12pt}
\includegraphics[width=0.61\textwidth]{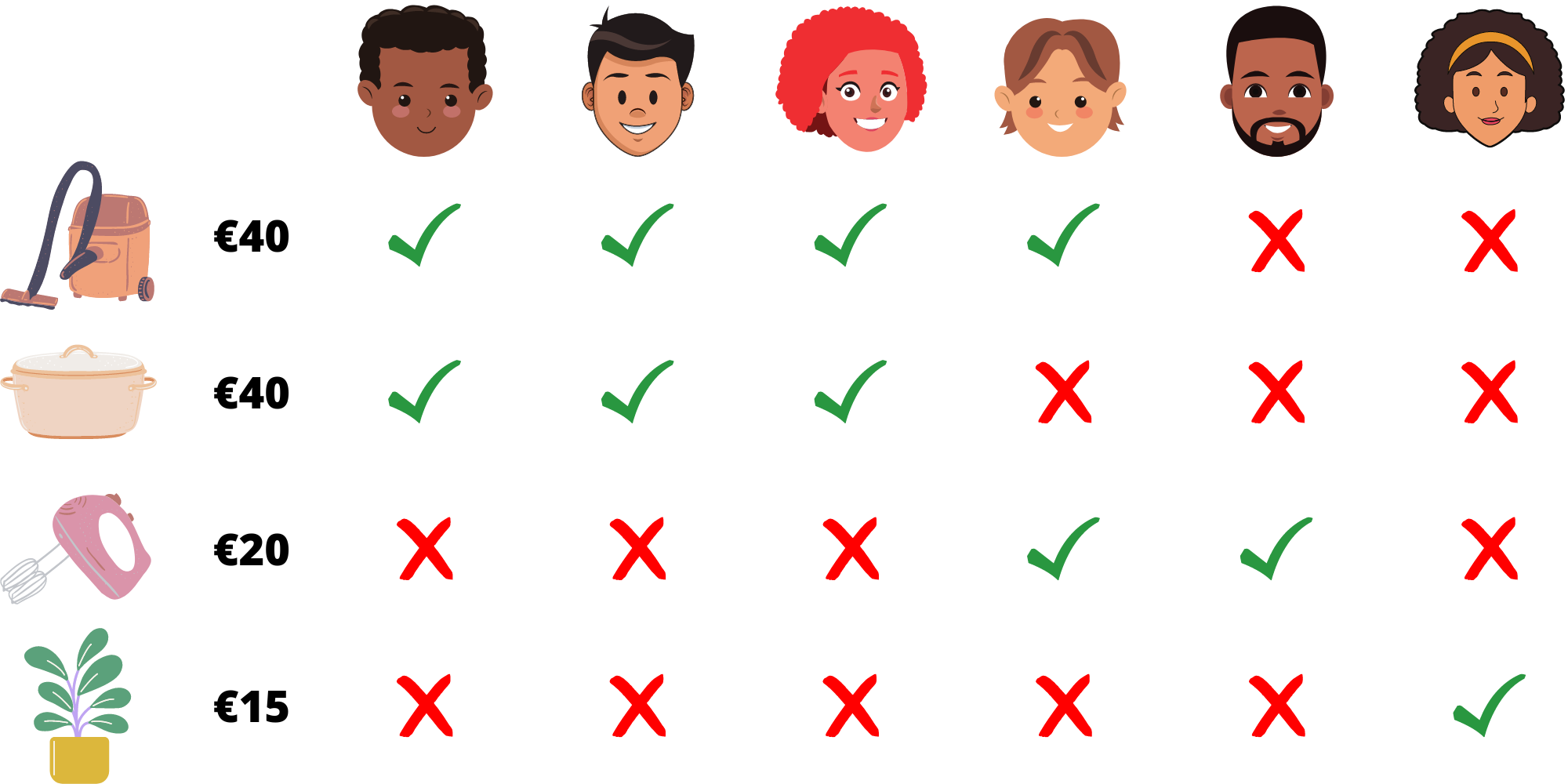}
\end{figure}

De \emph{Greedy Cost Welfare} (\rl{GreedCost}) regel kiest de items met de meeste stemmen: de stofzuiger en de pan.

\begin{figure}[h!]
\centering
\vspace{15pt}
\includegraphics[width=0.61\textwidth]{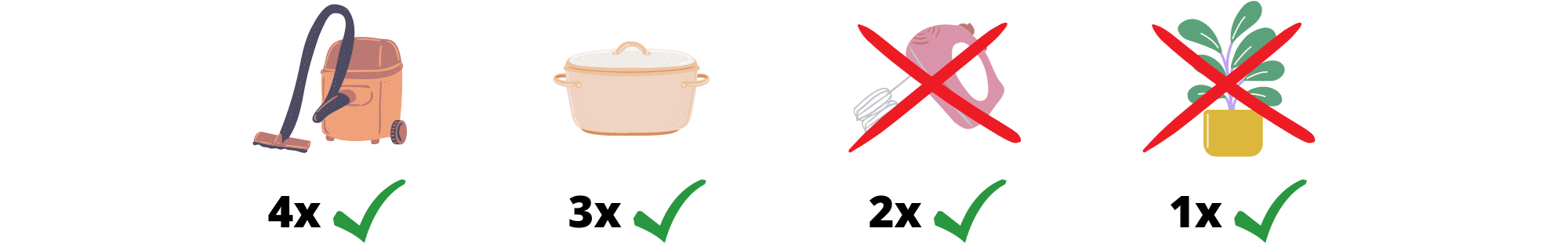}
\end{figure}

De \emph{Methode van Gelijke Delen} (\rl{MES}) verdeelt het budget eerlijk over alle stemmers en simuleert coalities die de items kopen. De kosten worden zo gelijkmatig mogelijk verdeeld. \rl{MES} selecteert de stofzuiger, de mixer en de plant. Daardoor blijft geen enkele stemmer met lege handen achter.

\begin{figure}[h!]
\centering
\vspace{12pt}
\includegraphics[width=0.61\textwidth]{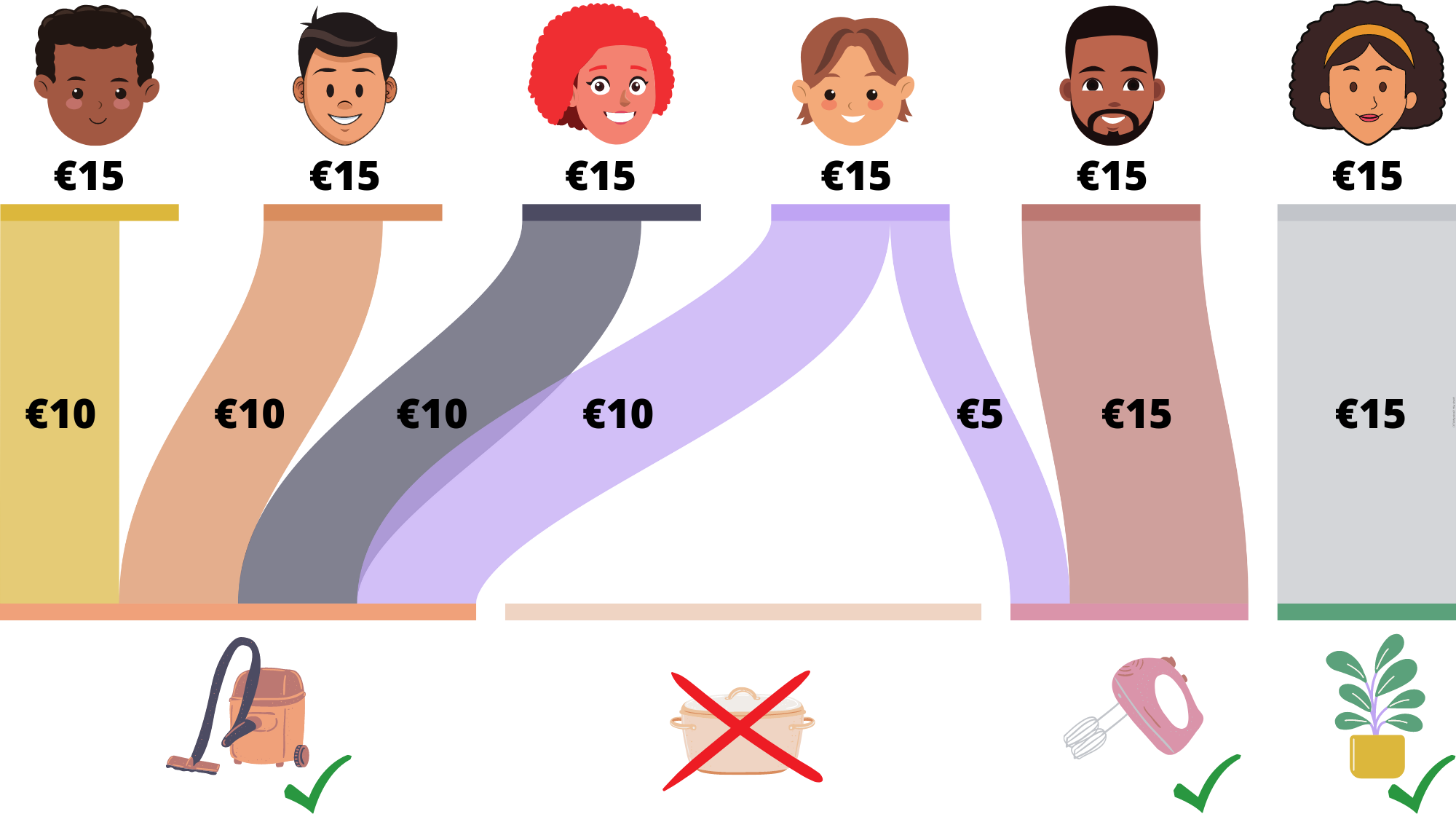}
\end{figure}

\subsection{Tekstuele uitleg}

\begin{otherlanguage}{dutch}
Het doel van Participatieve Begroting (PB) is om democratisch te beslissen welke projecten een overheid moet financieren en welke projecten zij moet afwijzen. Een PB proces bestaat uit het verzamelen van mogelijke projecten, het organiseren van een verkiezing en een bekendmaking van welke projecten gefinancierd worden. Tussen de laatste twee stappen worden de stemmen geteld en wordt er bepaald welke projecten winnen. Dat kan op meerdere manieren, waaronder de `Methode van Gelijke Delen', ofwel `Method of Equal Shares' (MES). Hieronder leggen we uit hoe MES werkt. Een andere uitleg staat op de website \url{https://equalshares.net/explanation}.

Het doel van MES is om het budget zo eerlijk mogelijk te verdelen; elke kiezer moet even veel invloed hebben op de uitslag. Dat doet deze stemregel door het budget virtueel te verdelen over alle kiezers, en de kiezers samen projecten te laten `kopen'. Wanneer de virtuele portemonnee van een kiezer leeg is, heeft de kiezer geen invloed meer op welke projecten nog meer gefinancierd worden.

Het stemmen gebeurt met goedkeurings-biljetten. Dat betekent dat elke kiezer op diens stembiljet aangeeft welke projecten die goedkeurt en welke projecten die afkeurt. Vervolgens wordt de MES-regel gebruikt om de stemmen te tellen.

MES kan gezien worden als een simulatie van kiezers die coalities vormen om samen projecten te `kopen'. De simulatie bestaat uit de volgende stappen.
\begin{enumerate}
    \item Het totale budget wordt eerlijk verdeeld over alle kiezers. De kiezers hebben dus een virtuele portemonnee met elk even veel geld.
    \item Voor elk project wordt gekeken of alle kiezers die dit project goedkeuren, samen genoeg geld in hun virtuele portemonnees hebben om het project te kopen. Is dat niet zo, dan wordt het project niet gefinancierd.
    \item Als er meerdere projecten overblijven, wordt er per project gekeken hoe \emph{eerlijk} de kosten van het project verdeeld kunnen worden onder de kiezers die het project goedkeuren. Het project waarbij dat het meest eerlijk kan, wordt door de kiezers gekocht. De kosten worden uit de virtuele portemonnees gehaald en de simulatie gaat terug naar stap 2.
    \item Wanneer elk project is afgekeurd of gekocht, eindigt de simulatie.
\end{enumerate}

Hoe \emph{eerlijk} de kosten van een project verdeeld kunnen worden, wordt als volgt bepaald.

Aanvankelijk worden de kosten volledig gelijk verdeeld over de kiezers die het project goedkeuren. Het zou echter kunnen dat sommige kiezers al andere projecten `gekocht' hebben en dus niet genoeg geld over hebben om dit gelijke deel te betalen. Als dat het geval is, betalen deze kiezers hun gehele overgebleven budgetten. De kosten die zij niet kunnen dekken, worden gelijk verdeeld over de andere kiezers die het project steunen en nog wel genoeg geld in hun portemonnee hebben.

De verdeling van de kosten wordt zo dus enigszins ongelijk. Hoe ongelijk de verdeling is, wordt berekend door de bijdrage die een kiezer moet leveren, $\gamma$, te delen door de totale kosten van het project, $c$. Die waarde is dus ${\alpha = \frac{\gamma}{c}}$. Onder de aanname dat de totale kosten van een project gelijk zijn aan de voldoening die het project oplevert, zijn $\alpha$ dus de betaalde kosten per eenheid voldoening. Wanneer de kosten volledig gelijk verdeeld kunnen worden, is $\alpha$ laag voor alle kiezers die het project ondersteunen. Wanneer de kosten zeer ongelijk verdeeld worden, is de $\alpha$ van de grootste bijdrager veel groter dan die van kleinere bijdragers. De kostenverdeling van het project met de laagste maximale~$\alpha$ wordt daarom als het meest \emph{eerlijk} gezien.

Al met al simuleert MES wat kiezers zouden doen, als we hun inderdaad allemaal een gelijk deel van het budget zouden geven en ze vrij zouden zijn om (zo eerlijk mogelijke) coalities te vormen om projecten te realiseren. Zo geeft de regel elke kiezer even veel invloed en bereikt MES een meer proportionele en gelijke verdeling van het budget.
\end{otherlanguage}

\newpage\section{Projects (with Category and Cost) Selected in Instance 522}
\label{app:522}

\begin{itemize}
\item Selected by both rules (€63.300):
\begin{itemize}
\item A greener and more pleasant Robijn Square! (Streets, greenery \& squares, €30.000)
\item Smaragd Street: Meeting place at the entrance of the Diamond neighborhood (Streets, greenery \& squares, €20.000)
\item An Iftar, a neighbourhood party and a Christmas meal (Doing things together, €7.000)
\item First aid for cardiac arrest - an AED can save lives in the neighbourhood (Health, well-being \& opportunities, €6.300)
\end{itemize}
\item Selected by \rl{GreedCost} only (€29.000):
\begin{itemize}
\item Container gardens against litter and more greenery (Streets, greenery \& squares, €29.000)
\end{itemize}
\item Selected by $\rl{mes}^{{*}{+}}$ only (€28.250):
\begin{itemize}
\item Replacement of playgrounds at Smaragd Square/Street (Streets, greenery \& squares, €20.000)
\item Musical gems of the Diamond neighbourhood (Doing things together, €8.250)
\end{itemize}
\end{itemize}

\section{Projects (with Category and Cost) Selected in Instance 613}
\label{app:613}

\begin{itemize}
\item Selected by both rules (€201.678):
\begin{itemize}
\item Trips and activities for children (Youth activities, €13.139)
\item Homework guidance (Equal opportunities \& education, €9.115)
\item Sewing lessons to combat loneliness and debt (Meeting \& connection, €9.600)
\item The eyes and ears of and in our neighbourhoods (Safety, €9.780)
\item Shade in the pasture for livestock/livability for animals (Meeting \& connection, €4.600)
\item Girls Kick-Off (Sports \& exercise, €11.638)
\item 30 worm hotels/facade benches on sidewalks of North-East (Meeting \& connection, €25.350)
\item Safety and greenery... you have to do it! (Safety, €14.000)
\item Well-prepared for the transition to secondary education (Equal opportunities \& education, €31.650)
\item Atelier Nieuwendam (Equal opportunities \& education, €4.120)
\item Addressing loitering youth in the neighbourhood (Youth activities, €32.926)
\item Sports and exercise for girls and women (Sports \& exercise, €17.300)
\item How do we promote equal opportunities in education? (Equal opportunities \& education, €8.880)
\item Soup at the Waaltje (Meeting \& connection, €9.580)
\end{itemize}
\item Selected by \rl{GreedCost} only (€59.000):
\begin{itemize}
\item Lower garden Plan van Gool (Meeting \& connection, €59.000)
\end{itemize}
\item Selected by $\rl{mes}^{{*}{+}}$ only (€57.781):
\begin{itemize}
\item The 5 World Countries Fashion Show (Meeting \& connection, €3.298)
\item Sports opportunities for immigrant women (Sports \& exercise, €13.075)
\item Flower power (Meeting \& connection, €3.104)
\item Go for a colourful Zunderdorp! (Meeting \& connection, €6.055)
\item Walking buddies (Sports \& exercise, €6.340)
\item Neighbourhood party with a food truck, DJ and an MC (Meeting \& connection, €7.325)
\item Sports and a board in Naardermeerstraat (Sports \& exercise, €2.260)
\item Sports and a board in de Kleine Wereld (Sports \& exercise, €2.160)
\item The Bea Creas (Equal opportunities \& education, €5.404)
\item Sports and a board in Plan van Gool (Sports \& exercise, €2.260)
\item Music and culture in nature for the neighborhood (Art \& creativity, €6.500)
\end{itemize}
\end{itemize}

\section{Projects (with Category and Cost) Selected in Instance 644}
\label{app:644}

\begin{itemize}
\item Selected by both rules (€118.500):
\begin{itemize}
\item Play park for all ages (Public space, €60.000)
\item Repair shop for walkers and mobility scooters (Other, €5.000)
\item Community dinner for the homeless (Social, €7.000)
\item Everything market (Public space, €11.500)
\item Container gardens Noordzijde (Public space, €30.000)
\item Benches in the Tuinen van West nature reserve (Green, €5.000)
\end{itemize}
\item Selected by \rl{GreedCost} only (€147.000):
\begin{itemize}
\item Water fountains/trick fountains at Lambertus Zijlplein (Public space, €75.000)
\item Makeover of Confucius playground (Public space, €72.000)
\end{itemize}
\item Selected by $\rl{mes}^{{*}{+}}$ only (€147.260):
\begin{itemize}
\item Ping pong in Noorderhof (Public space, €5.000)
\item Empower Women: training and sports lessons (Social, €7.700)
\item Keep New West Clean: speak out and make your street your second home (Culture, €15.500)
\item Trunk market in Geuzenveld, Slotermeer (Culture, €6.000)
\item Girls of New West in their power (Social, €6.700)
\item Painting concrete blocks at Jan de Jonghkade (Public space, €5.000)
\item Karaoke New West (Social, €7.500)
\item Sports facilities for and by young people (Social, €9.400)
\item A greenhouse to give away plants in the Volkstuinpark de Bretten (Green, €10.000)
\item Recycled tanks for rainwater harvesting (Green, €7.000)
\item Addressing learning deficiencies caused by COVID-19 (Social, €23.000)
\item Kumbet Foundation against elderly loneliness (Social, €30.000)
\item Less conflict and more connection through workshops for young people (Other, €6.000)
\item Activities for lonely women in New West (Social, €8.460)
\end{itemize}
\end{itemize}

\end{document}